\documentclass{article}
\usepackage{graphicx}

\def\abstract#1{\vskip 7mm 
        \begin{center}{\large Abstract}\par \smallskip
                \begin{minipage}[c]{12cm}
                        \small #1
                \end{minipage}
        \end{center}
}
\def\title#1{\begin{center}{\Large\bf #1}\end{center}}
\def\author#1{\vskip 5mm \begin{center}{#1}\end{center}}
\def\address#1{\begin{center}{\it #1}\end{center}}
\def\be{\begin{equation}}
\def\ee{\end{equation}}
\newcommand{\bea}{\begin{eqnarray}}
\newcommand{\eea}{\end{eqnarray}}
\newcommand{\vx}{\ensuremath{\vec{x}}}
\newcommand{\vk}{\ensuremath{\vec{k}}}
\makeatletter
\@ifundefined{lesssim}{\def\lesssim{\mathrel{\mathpalette\vereq<}}}{}
\@ifundefined{gtrsim}{}{}
\def\vereq#1#2{\lower3pt\vbox{\baselineskip1.5pt \lineskip1.5pt
\ialign{$\m@th#1\hfill##\hfil$\crcr#2\crcr\sim\crcr}}}
\makeatother

\begin{document}

\title{%
  Clarifying Slow Roll Inflation and the Quantum Corrections to the Observable
Power Spectra
}
\author{%
D. Boyanovsky$^{c,a,}$\footnote{E-mail:boyan@pitt.edu},
\underline{H. J. de Vega}$^{b,a,}$\footnote{E-mail:devega@lpthe.jussieu.fr},
N. G. Sanchez $^{a,}$\footnote{E-mail:Norma.Sanchez@obspm.fr}
}
\address{%
$^{a}$
Observatoire de Paris, LERMA, Laboratoire Associ\'e au CNRS UMR 8112,
\\ 61, Avenue de l'Observatoire, 75014 Paris, France. 
\\$^{b}$ LPTHE, Laboratoire Associ\'e au CNRS UMR 7589,
Universit\'e Pierre et Marie Curie (Paris VI) \\ et Denis Diderot (Paris VII),
Tour 24, 5 \`eme. \'etage, 4, Place Jussieu, 75252 Paris, Cedex 05,
France. \\
$^{c}$Department of Physics and
Astronomy, University of Pittsburgh, Pittsburgh, Pennsylvania 15260,
USA }
\abstract{Slow-roll inflation can be studied as an effective field theory.
The form of the inflaton potential consistent with the data is 
$ V(\phi) = N \; M^4 \; w\left(\frac{\phi}{\sqrt{N} \; M_{Pl}}\right) $ 
where $ \phi $ is the inflaton field, $ M $ is the inflation energy scale,   
and $ N \sim 50 $ is the number of efolds since the cosmologically
relevant modes exited the Hubble radius until the end of
inflation. The dimensionless function $ w(\chi) $ and field
$ \chi $ are generically $ \mathcal{O}(1) $. The WMAP value for
the amplitude of scalar adiabatic fluctuations yields $ M \sim
0.77\times 10^{16}$GeV. This form of the potential encodes 
the slow-roll expansion as an expansion in $1/N$.  A
Ginzburg-Landau (polynomial) realization of $ w(\chi) $
reveals that the Hubble parameter, inflaton mass and non-linear
couplings are of the see-saw form in terms of the small ratio $
M/M_{Pl} $. The quartic coupling is $ \lambda \sim
\frac{1}{N} \left(\frac{M}{M_{Pl}}\right)^4 $. The smallness of
the non-linear couplings is {\bf not} a result of fine tuning but
a {\bf natural} consequence of the validity of the effective field
theory and slow roll approximation. Our observations suggest that slow-roll
inflation may well be described by an almost critical
theory, near an infrared stable gaussian fixed point.
Quantum corrections to slow roll inflation are computed and turn to
be an expansion in  powers $ \left(H/M_{Pl}\right)^2 $. 
The corrections to the inflaton effective potential and its equation of motion 
are computed, as well as the quantum corrections to the observable
power spectra. The near scale invariance of the fluctuations
introduces a strong infrared behavior naturally regularized by the
slow roll parameter $ \Delta \equiv \eta_V-\epsilon_V = \frac12(n_s -1)+
r/8 $. We find the \emph{effective} inflaton potential during slow roll
inflation including the contributions from scalar curvature and tensor
perturbations as well as from light scalars and Dirac
fermions coupled to the inflaton. The scalar and tensor
superhorizon contributions feature infrared enhancements regulated by
slow roll parameters. Fermions and gravitons
do not exhibit infrared enhancement. The subhorizon part is
completely specified by the trace anomaly of the fields with
different spins and is solely determined by the space-time geometry.
This inflationary effective potential is strikingly {\bf different} 
from the usual Minkowski space-time result. 
Quantum corrections to the power spectra are expressed in terms of the 
CMB observables: $ n_s, \; r $ and $ dn_s/d \ln k $. 
Trace anomalies (especially the
graviton part) dominate these quantum corrections in a definite
direction: they {\bf enhance} the scalar curvature fluctuations and
{\bf reduce} the tensor fluctuations. }

\section{Inflation as an Effective Field Theory}

Inflation was originally proposed to solve several outstanding
problems of the standard Big Bang model
\cite{guthsato} thus becoming an important
paradigm in cosmology. At the same time, it provides a natural
mechanism for the generation of scalar density fluctuations that
seed large scale structure, thus explaining the origin of the
temperature anisotropies in the cosmic microwave background (CMB),
as well as that of  tensor perturbations (primordial gravitational
waves).  A distinct aspect of
inflationary perturbations is that these are generated by quantum
fluctuations of the scalar field(s) that drive inflation. 
Their physical wavelengths grow faster than the Hubble radius and when
they cross the horizon they freeze out and decouple.
Later on, these fluctuations are amplified and grow, becoming classical and
decoupling from causal microphysical processes. Upon re-entering
the horizon, during the matter era, these  scalar (curvature)
perturbations induce temperature anisotropies imprinted
on the CMB at the last scattering surface and 
seed the inhomogeneities which generate structure upon gravitational
collapse\cite{pert,hu}.  Generic
inflationary models predict that these fluctuations are adiabatic
with an almost scale invariant spectrum. Moreover, they are  Gaussian 
to a very good approximation. 
These generic predictions are in spectacular agreement with the CMB
observations as well as with a variety of large scale structure data
\cite{WMAP}. The WMAP data \cite{WMAP} clearly display an anti-correlation
peak in the temperature-polarization (TE) angular power spectra at
$l\sim 150$, providing one of the most striking confirmations of superhorizon
adiabatic fluctuations as predicted by inflation\cite{WMAP}. 

The robust predictions of inflation (value of the  entropy of the universe, 
solution of the flatness problem, small adiabatic Gaussian density 
fluctuations explaining the CMB anisotropies, ...) which are common to the
available inflationary scenarios, show the predictive power of the 
inflationary paradigm. While there is a great diversity of inflationary models,
they predict fairly generic features: a gaussian, 
nearly scale invariant spectrum of
(mostly) adiabatic scalar and tensor primordial fluctuations. 
More precisely, the WMAP\cite{WMAP} data can be fit outstandingly well 
by simple single field slow roll models.
These generic predictions of  inflationary models make the
inflationary paradigm  robust. Whatever the microscopic model for the early 
universe (GUT theory) would be, it should include inflation with the generic 
features we know today. 

Inflationary dynamics is typically studied by treating  the
inflaton as a homogeneous  classical scalar
field whose evolution is determined
by its classical equation of motion, while the inflaton 
quantum fluctuations (around the classical value and in the
Gaussian approximation) provide the seeds for the scalar
density perturbations of the metric. The classical dynamics of the 
inflaton (a massive scalar field) 
coupled to a cosmological background clearly shows that inflationary
behaviour is an {\bf attractor} \cite{bgzk}. This is a generic and robust
feature of inflation. 

In quantum field theory,
this classical inflaton corresponds to the expectation value 
of a quantum field operator in a translational invariant state.
Important aspects of the inflationary dynamics, as resonant
particle production and the nonlinear back-reaction that it generates,
require a full quantum treatment of the inflaton for their consistent
description. The quantum dynamics of the inflaton 
in a non-perturbative framework and its consequences on the CMB anisotropy 
spectrum were treated in refs.\cite{cosmo,cosmo2,staro,anom}. 

The quantum fluctuations are of two different kinds: (a) Large quantum 
fluctuations generated at the begining of inflation through
spinodal or parametric resonance depending on the inflationary scenario
chosen. They have comoving wavenumbers in the range of $ 10^{13} \mbox{GeV} 
\lesssim  k \lesssim 10^{15}$GeV and they become superhorizon a few efolds
after the begining of inflation. Their physical wavenumbers
become  subsequently very small compared with the inflaton mass.
Therefore, the assembly of these modes can be treated as part of the zero mode
after $5-10$ efolds \cite{cosmo,cosmo2}. That is, the use of an homogeneous
classical inflaton is thus justified by the full quantum theory treatment of 
the inflaton.
(b) Small fluctuations of high comoving wavenumbers 
$$
e^{N_T-60} \; 10^{16} \, GeV < k < e^{N_T-60} \; 10^{20} \, GeV
$$
where $ N_T \geq 60 $ stands for the total number of efolds (see for example
Ref. \cite{sd}). These are the cosmologically relevant modes that exit
the horizon about 50 efolds before the end of inflation and
reenter later on (during the matter dominated era) being the source of
primordial power for the CMB anisotropies as well as for the structure
formation. While modes (b) obey linear evolution equations with
great accuracy, modes (a) strongly interact with themselves calling
for nonperturbative quantum field theory treatments as in 
refs.\cite{cosmo,cosmo2}. Notice that particle production is governed by
linear unstabilities (parametric or spinodal) only at the begining
of inflation. Particle production keeps strong during the nonlinear
regime till particles eventually dominate the energy density and inflation 
stops\cite{cosmo}. The modes (b) correspond to physical scales that were microscopic 
(even transplanckian) at the
begining of inflation, then after they become astronomical and produce  the CMB
anisotropies as well as the large scale structure of the universe. 

\begin{figure}[h]
\includegraphics[width=12cm,height=8cm]{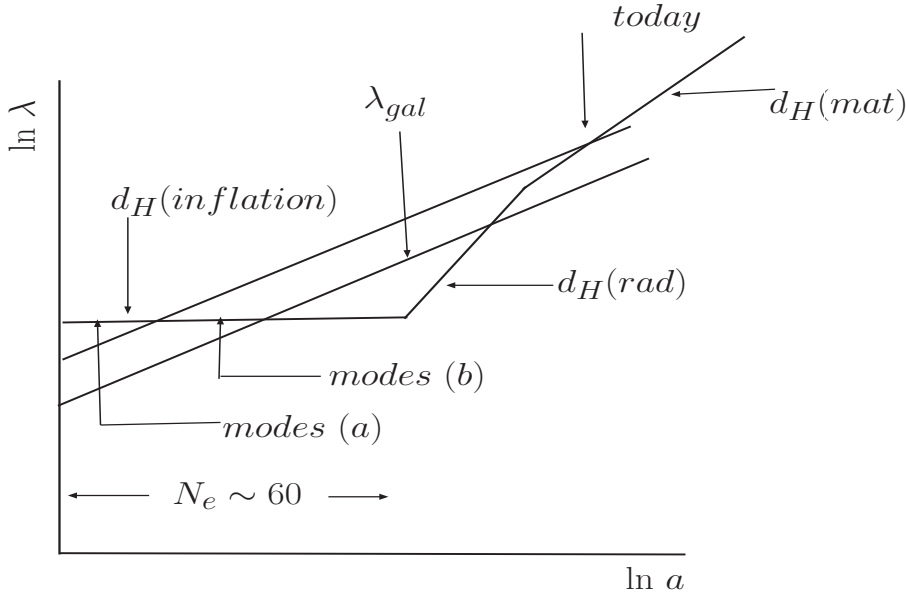}
\caption{ Physical lengths $ \lambda = a(t) \;
\lambda_{comoving} $ vs. the scale factor $  a(t) $ in a log-log plot. 
The causal horizon $ d_H $ is shown for the inflationary
(de~Sitter), radiation dominated and matter dominated eras. The
physical wavelengths ($ \lambda $) for today's Hubble radius
and a typical galactic scale ($ \lambda_{gal} $) are shown. One sees the modes (b)
  can cross the horizon {\bf twice} bringing information from
  extremely short wavelength modes during the
  inflationary era. Modes (a) which have large amplitudes during inflation
and dominated the energy of the universe, have not yet crossed back inside
the horizon.}
\label{infl}
\end{figure}

The crucial fact is that the excitations can
cross the horizon {\bf twice}, coming back and bringing information
from the inflationary era. We depict in fig. \ref{infl} the physical 
wavelengths of modes (a) and (b) vs. the logarithm of the scale factor
showing that modes (b) crossed twice the horizon, modes (a) are 
out of the horizon still today.

Recently, particle decay in a de Sitter background as well as  during slow
roll inflation has been studied in ref.\cite{pardec} together with
its implication for the decay of the density fluctuations. Quantum
effects during slow roll inflation including quantum corrections
to the effective inflaton potential and its equation of motion are
derived in ref.\cite{anom,qua}.
Recent studies of quantum corrections during
inflation\cite{anom,pardec,qua} revealed the robustness of classical
single field slow roll inflationary models as a result of the
validity of the {\bf effective field theory} description. The reliability
of an effective field theory of inflation hinges on a wide
separation between the energy scale of inflation, determined by $H$
and that of the underlying microscopic theory which is taken to be
the Planck scale $ M_{Pl} $.
Inflation as known today should be considered as an {\bf effective theory},
that is, it is not a fundamental theory but a theory of a
condensate (the inflaton field) which follows from a more fundamental one 
(the GUT model). The inflaton field is just an {\bf effective}
description while the microscopic description should come from the GUT model
in the cosmological spacetime. Such derivation is not yet available.

Bosonic fields do
not need to be fundamental fields, for example they may emerge as
condensates of fermion-antifermion pairs $ < {\bar \Psi} \Psi> $
in a grand unified theory (GUT) in the cosmological background. In
order to describe the cosmological evolution is enough to consider
the effective dynamics of such condensates.  The relation between
the low energy effective field theory of inflation and the
microscopic fundamental theory is akin to the relation between the
effective Ginzburg-Landau theory of superconductivity and the
microscopic BCS theory, or like the relation of the $ O(4) $ sigma
model, an effective low energy theory of pions, photons and chiral
condensates with quantum chromodynamics (QCD)\cite{quir}. The
guiding principle to construct the effective theory is to include
the appropriate symmetries\cite{quir}. Contrary to the sigma model
where the chiral symmetry strongly constraints the
model\cite{quir}, only general covariance can be imposed on the
inflaton model.

The inflationary scenarios are usually
characterized as small and large fields scenarii. In small fields
scenarios the initial classical amplitude of the inflaton is assumed
small compared with $ M_{Pl} $ while in large field scenarii
the inflaton amplitude is initially of the order $ M_{Pl} $. The 
first type of scenarii is usually realized with broken symmetric
potentials ($ m^2 < 0 $) while for the second type scenarii (`chaotic
inflation') one can just use unbroken potentials with $ m^2 > 0 $.

Gravity can be treated semiclassically for inflation: the geometry is
classical and the metric follows from the Einstein-Friedman equations
where the r.h.s. is the expectation value of a quantum
operator. Quantum gravity corrections can be neglected during
inflation because the energy scale of inflation $ M \sim M_{GUT} \sim 10^{-3} 
\; M_{Planck} $. That is, quantum gravity
effects are at most $ (M/M_{Planck})^2 \sim 10^{-6} $ and can be neglected in this
context. The studies in ref.\cite{pardec,qua,clar} reveal
that  quantum corrections in the effective field theory yields an
expansion in $ \left(\frac{H}{ M_{Pl}}\right)^2 $ for
\emph{general inflaton potentials}. This indicates
that the use of the inflaton potential $ V(\phi) $ from effective
field theory is consistent for
$$
 \left(\frac{H}{M_{Pl}}\right)^2 \ll 1 \quad {\rm and~hence} \quad
V(\phi) \ll  M_{Pl}^4 \; ,
$$
allowing amplitudes of the inflaton field $ \phi $ 
{\bf well beyond} $ M_{Pl} $\cite{clar}. 

\subsection{Slow-roll Inflation as an expansion in $1/N_{efolds}$ and
no fine tuning}

In single field inflation the energy density is dominated by a
homogeneous scalar \emph{condensate}, the inflaton, whose dynamics
can be described by an  \emph{effective} Lagrangian
\be\label{lagra}
{\cal L} = a^3({ t}) \left[ \frac{{\dot
\phi}^2}{2} - \frac{({\nabla \phi})^2}{2 \;  a^2({ t})} - V(\phi) \right]
\; ,
\ee
together to  the Einstein-Friedman equation
\be \label{ef}
\left[ \frac{1}{a(t)} \; \frac{da}{d t} \right]^2 =
\frac{\rho( t)}{3 \; M^2_{Pl}}   \;  ,
\ee
where the  energy density
for an homogeneous inflaton is given by
$$
\rho(t) = \frac{{\dot \phi}^2}{2} +  V(\phi) \; .
$$
\noindent The inflaton potential $ V(\phi) $ is a slowly varying
function of $ \phi $ in order to permit a slow-roll solution for
the inflaton field $ \phi(t) $. All this in a spatially flat FRW metric
\be \label{metrica}
ds^2= dt^2-a^2(t) \; d{\vec x}^2 = C^2(\eta) \left[ (d \eta)^2 -  d{\vec x}^2
\right] \; ,
\ee
where $ \eta $ is the  conformal time and $ C(\eta) \equiv a(t(\eta)) $.

The inflaton evolution equation takes the form,
$$
{\ddot \phi} + 3 \, H \, {\dot \phi} +  V'(\phi) = 0 \; .
$$
In the slow-roll approximation $ {\ddot \phi} \ll  
3 \, H \, {\dot \phi} $ and $ \frac{{\dot \phi}^2}{2} \ll  V(\phi) $,
and the inflaton evolution equations become
\be \label{slow}
 3 \, H(t) \, {\dot \phi} +  V'(\phi) = 0 \quad , \quad
\left[ H(t) \right]^2 = \frac{ V(\phi)}{3 \; M^2_{Pl}}  \; ,
\ee
where $ H(t) \equiv \frac{1}{a(t)} \; \frac{da}{dt} $ stands for the Hubble
parameter. Eq.(\ref{slow}) can be integrated by quadratures
with the result 
$$
\ln a(t) = - \frac1{M^2_{Pl}} \int^{\phi(t)}_{\phi(0)} \frac{V(\phi)}{V'(\phi)}
\; d\phi \; .
$$
This formula shows that the inflaton field $ \phi $ scales as $ M_{Pl} $
and as the square root of the number of efolds \cite{clar}.
This suggest to introduce the
dimensionless field $ \chi $ and the dimensionless potential $ w(\chi) $,
\be \label{defxi}
\chi \equiv  \frac{\phi}{\sqrt{N} \; M_{Pl}} 
\qquad , \qquad  w(\chi) \equiv \frac{V(\phi)}{ N \; M^4 }
\ee
where $ M $ stands for the scale of inflation.
The dimensionless field $ \chi $ is \emph{slowly} varying during the stage
of slow roll inflation: a large change in the field amplitude $\phi$
results in a small change in $\chi$ amplitude ,
\be \label{slofield}
\Delta \chi = \frac{1}{\sqrt{N}} \frac{\Delta \phi}{M_{Pl}}  \; ,
\ee
a change in $\phi$ with $\Delta \phi \sim
M_{Pl}$ results in a change $\Delta \chi \sim 1/\sqrt{N}$.

Introducing a \emph{stretched}  (slow) dimensionless time variable $\tau$ and a
rescaled dimensionless Hubble parameter $h$ as follows
\be\label{time} 
t = \sqrt{N} \; \frac{M_{Pl}}{M^2} \;  \tau \qquad
, \qquad H = \sqrt{N} \;  \frac{M^2}{M_{Pl}}\; h \; , 
\ee 
the Einstein-Friedman equation reads 
\be \label{efa} 
h^2(\tau) = \frac13\left[\frac1{2\;N} \left(\frac{d\chi}{d \tau}\right)^2 +
w(\chi) \right] \; , 
\ee 
and the evolution equation for the inflaton field
$\chi$ is given by 
\be \label{eqnmot} 
\frac1{N} \;  \frac{d^2
\chi}{d \tau^2} + 3 \; h \; \frac{d\chi}{d \tau} + w'(\chi) = 0 \; .
\ee 
The slow-roll approximation follows by neglecting the
$\frac1{N}$ terms in eqs.(\ref{efa}) and (\ref{eqnmot}). Both
$w(\chi)$ and $h(\tau)$ are of order $N^0$ for large $N$. Both
equations make manifest the slow roll expansion as a {\bf systematic expansion in}
$1/N$\cite{clar}. 

Following the spirit of the Ginsburg-Landau theory of phase transitions,
the simplest choice is a quartic trinomial for the inflaton potential\cite{nos}:
\be \label{wxi}
w(\chi)= w_0 \pm \frac12 \; \chi^2 + \frac{G_3}3 \; \chi^3 +
\frac{G_4}{4} \; \chi^4 \; .
\ee
where the coefficients $ w_0, \; G_3 $ and $ G_4 $ are of order one
and the signs $ \pm $ correspond to large and small field inflation, 
respectively. Inserting eq.(\ref{wxi}) in eq.(\ref{defxi}) yields,
\be\label{VI}
V(\phi)= V_0 \pm \frac{m^2}{2} \; \phi^2 +  \frac{ m
\; g }{3} \; \phi^3 + \frac{\lambda}{4}\; \phi^4 \; .
\ee
where $ m², \; g $ and $ \lambda $ are given by  the
following see-saw-like relations, 
\be 
m = \frac{M^2}{M_{Pl}} \qquad ,  \qquad 
g = \frac{G_3}{\sqrt{N}} \left( \frac{M}{M_{Pl}}\right)^2  
\qquad ,  \qquad 
\lambda  = \frac{G_4}{{N}} \left( \frac{M}{M_{Pl}}\right)^4 
\label{aco} \; . 
\ee 
We can now input the results from WMAP\cite{WMAP} to constrain the
scale $M$. The amplitude of adiabatic scalar perturbations in
slow-roll is expressed as
\be \label{ampliI}
|{\Delta}_{k\;ad}^{(S)}|^2  = \frac{1}{12 \, \pi^2 \;  M_{Pl}^6}
\; \frac{V^3}{V'^2}= \frac{N^2}{12 \, \pi^2} \;
\left(\frac{M}{M_{Pl}}\right)^4 \; \frac{w^3(\chi)}{w'^2(\chi)} \; ,
\ee
Since, $ w(\chi) $ and  $ w'(\chi) $ are of order one, we find
\be\label{Mwmap}
\left(\frac{M}{M_{Pl}}\right)^2 \sim \frac{2
\, \sqrt{3} \, \pi}{N} \; |{\Delta}_{k\;ad}^{(S)}| \simeq  1.02
\times 10^{-5} \; .
\ee
where we used $ N \simeq 50 $ and the WMAP
value for $ |{\Delta}_{k\;ad}^{(S)}| = (4.67 \pm 0.27)\times
10^{-5} $ \cite{WMAP}. This fixes the scale of inflation to be
$$
M \simeq 3.19 \times 10^{-3} \; M_{PL} \simeq 0.77
\times 10^{16}\,\textrm{GeV} \; .
$$
This value pinpoints the scale of
the potential during inflation to be at the GUT scale suggesting a
deep connection between inflation and the physics at the GUT
scale in cosmological space-time.

That is, the WMAP data {\bf fix} the scale of inflation $ M $ for
single field potentials with the form given by eq.(\ref{defxi}).
This value for $M$ is below the WMAP upper bound on the inflation scale
$ 3.3 \; 10^{16}$GeV \cite{WMAP}.
Furthermore, the Hubble scale during (slow roll) inflation and the 
inflaton mass near the minimum of the potential are thereby determined from
eqs.(\ref{time}) and (\ref{aco}) to be $
 m = \frac{M^2}{M_{Pl}} = 2.45 \times 10^{13} \,\textrm{GeV},
\quad  H  = \sqrt{N} \; m \; h \sim 1.0 \times 10^{14}\,\textrm{GeV}
= 4.1 \; m \;  $ since $ h = {\cal O}(1) $.
In addition, the order of magnitude of the couplings naturally follows
from eq.(\ref{aco}): $ g \sim 10^{-6}, \quad \lambda \sim 10^{-12}$, 
since $ M/M_{Pl} \sim 3 \times 10^{-3} $.

Since $M/M_{Pl} \sim 3 \times 10^{-3}$, 
these relations are a {\bf natural} consequence
of the validity of the effective field theory and of slow roll and
relieve the {\bf fine tuning problem}.  We emphasize that the
`see-saw-like' form of the couplings is a natural consequence of
the form of the potential eq.(\ref{defxi}).
While the hierarchy between the Hubble parameter,
the inflaton mass and the Planck scale during slow roll inflation
is well known, our analysis reveals that small couplings are
naturally explained in terms of powers of the ratio between the
inflationary and Planck scales \emph{and} integer powers of $ 1/\sqrt{N} $.

This is one of the main results presented in this lecture:
the effective field theory and slow roll descriptions of inflation,
both validated by WMAP, lead us to conclude that there is \emph{no
fine tuning} problem\cite{clar}. The smallness of the inflaton mass and the
coupling constants in this trinomial realization of the
inflationary potential is a \emph{direct} consequence of the
validity of both the effective field theory and the slow roll
approximations through a see-saw-like mechanism.

It must be stressed that these order of magnitude estimates follow
from the statement that  $ w(\chi) $ and  $ \chi $ are of order one.
They are thus independent of the details of the model.
Indeed, model-dependent factors of order one appearing in the
observables should allow to exclude or 
accept a given model by using the observational data.

The WMAP results rule out
the purely quartic potential ($ m=0, \; g=0 $). From the point of view
of an effective field theory this is not surprising:
it is rather \emph{unnatural} to set
$m=0$, since $m=0$ is a particular point at which the correlation
length is infinite and the theory is critical. Indeed the
systematic study in ref.\cite{nos} shows that the best fit to the
WMAP data requires $ m^2 \neq 0 $.

The general quartic Lagrangian eq.(\ref{V}) describes a
renormalizable theory. However, one can choose in the present context
arbitrary high order polynomials for  $ V(\phi) $.
These nonrenormalizable models
are also effective theories where $ M_{Pl} $ plays the r\^ole of UV cutoff.
However, already a quartic potential is rich enough to describe the full
physics and to reproduce accurately the WMAP data \cite{nos}.

For a general potential $ V(\phi) $,
\be \label{serie}
V(\phi) = \sum_{n=0}^{\infty} \lambda_n \; \phi^n \quad i. e.  \quad
 w(\chi) = \sum_{n=0}^{\infty} G_n \; \chi^n \; ,
\ee
we find from eq.(\ref{defxi}):
\be\label{Gn}
\lambda_n =  \frac{G_n \; m^2}{\left( N \; M_{Pl}^2 \right)^{\frac{n}2-1}}\; ,
\ee
where the dimensionless coefficients $ G_n $ are of order one.
We find the scaling behavior $\lambda_n \sim 1/N^{\frac{n}2-1}$.
Eq. (\ref{aco}) displays particular cases of eq.(\ref{Gn}) for $ n=3 $
and $ 4 $.

\medskip

The slow-roll parameters naturally result of the order $1/N, \; 1/N^2 $ etc.
when expressed in terms of the inflaton potential   
\be\label{VV}
V(\phi) =  N \; M^4 \; w(\chi) \; .
\ee
That is, 
\bea
&&\epsilon_V  = \frac{M^2_{Pl}}{2} \;
\left[\frac{V^{'}(\Phi_0)}{V(\Phi_0)} \right]^2  = \frac1{2 \; N} \;
\left[\frac{w'(\chi)}{w(\chi)} \right]^2 
\quad , \quad
\eta_V   = M^2_{Pl}  \; \frac{V^{''}(\Phi_0)}{V(\Phi_0)} =
\frac1{N}  \; \frac{w''(\chi)}{w(\chi)} \; , \label{etav}
\eea
The spectral index $ n_s $,  its running and the ratio of tensor to scalar 
fluctuations are expressed as
\bea \label{indi}
&&n_s - 1 = -\frac3{N} \; \left[\frac{w'(\chi)}{w(\chi)} \right]^2
+  \frac2{N}  \; \frac{w''(\chi)}{w(\chi)} \quad , \cr \cr 
&&\frac{d n_s}{d \ln k}= - \frac2{N^2} \; \frac{w'(\chi) \;
w'''(\chi)}{w^2(\chi)} - \frac6{N^2} \; \frac{[w'(\chi)]^4}{w^4(\chi)}
+ \frac8{N^2} \; \frac{[w'(\chi)]^2 \; w''(\chi)}{w^3(\chi)}\quad , \cr \cr 
&&r = \frac8{N} \; \left[\frac{w'(\chi)}{w(\chi)} \right]^2 \quad .
\eea
In eqs.(\ref{ampliI}), (\ref{etav}) and (\ref{indi}), the field $ \chi $ is computed 
at horizon exiting. We choose $ N[\chi] = N = 50 $.
The WMAP results favoured single inflaton models and among them new and hybrid
inflation emerge to be preferable than chaotic inflation\cite{nos}.

The inflationary era ends when the particles produced during inflation
dominate the energy density overcoming the vacuum energy.
At such stage the universe slows down its expansion to a radiation dominated
regime.

\subsection{Connection with Supersymmetry}

The form of the inflaton potential 
\be \label{V} 
V(\phi) = N \; M^4 \; w(\chi) 
\ee 
resembles (besides the factor $ N $) the moduli potential arising 
from supersymmetry breaking
\be\label{susy} 
V_{susy}(\phi) = m_{susy}^4 \; v\!\left(\frac{\phi}{M_{Pl}}\right) \; , 
\ee 
where $ m_{susy} $ stands for the supersymmetry breaking scale. In our
context, eq.(\ref{susy}) indicates  that $  m_{susy} \simeq M
\simeq M_{GUT} $. That is, the susy breaking scale $ m_{susy} $
turns out to be at the GUT and inflation scales. This may be a
\emph{first} observational indication of the presence of
supersymmetry. It should be noticed that this supersymmetry scale
is unrelated to an eventual existence of supersymmetry at the TeV scale.

Notice that the invariance of the basic interactions (the lagrangian)
and the invariance of the physical states (or density matrices) describing 
the matter are different issues.
For example, no thermal state at non-zero temperature
can be invariant under supersymmetry since Bose-Einstein and
Fermi-Dirac distributions are different. 
More generally, the inflationary stage is described by a scalar condensate
(the inflaton) while fermions cannot condense due to Pauli principle.
This makes quite hard to uncover supersymmetric properties of the lagrangian
from the physics of the early universe.

\subsection{Conjecture: inflation is near
a trivial infrared fixed point}\label{conjecture}
There are several remarkable features and consistency checks of the
relations (\ref{aco}):
\begin{itemize}
\item{Note the relation $\lambda \sim g^2$. This is the correct
consistency relation in a renormalizable theory because at one
loop level there is a renormalization of the quartic coupling (or
alternatively a contribution to the four points correlation
function) of orders $ \lambda^2 ,  \;  g^4 $ and $ \lambda \; g^2 $ which
are of the same order for $ \lambda \sim g^2 $. Similarly, at one
loop level there is a renormalization of the cubic coupling
(alternatively, a contribution to the three point function) of
orders $ g^3 $ and $ \lambda \; g $ which again require $ g^2 \sim
\lambda $ for consistency. }

\item {In terms of the effective field theory ratio $ (H/M_{Pl})^2 $ and
slow roll parameters, the dimensionless couplings are
\be
\frac{m \;  g}{H}  \sim  \frac1{N} \;
\frac{H}{M_{Pl}}  \qquad , \qquad
\lambda  \sim  \frac{1}{N^2} \, \left(\frac{H}{M_{Pl}} \right)^2
\label{lam} \; .
\ee
These relations agree with those found for the
dimensionless couplings in ref.\cite{pardec,qua} once the slow
roll parameters are identified with the expressions
(\ref{etav}) in terms of $1/N$.  The results of
refs. \cite{pardec,qua} revealed that the loop expansion is indeed an expansion
in the effective field theory ratio $ (H/M_{Pl})^2 $ and the slow roll
parameters. The study in ref.\cite{clar} allows us to go further in this direction
and state that
the loop expansion is a consistent double series in the effective
field theory ratio $ (H/M_{Pl})^2 $ \emph{and} in $ 1/N $. Loops are either
powers of $ g^2 $ or of $ \lambda $ which
implies that for each loop there is a factor $ (H/M_{Pl})^2 $. The
counting of powers of $ 1/N $ is more subtle: the nearly scale invariant
spectrum of fluctuations leads to  infrared enhancements of quantum
corrections in which the small factor $ 1/N $ enters as an infrared
regulator. Therefore, large denominators that feature the infrared
regulator of order $ 1/N $ cancel out factors $ 1/N $ in the
numerator. The final power of $ 1/N $  must be computed in
detail in each loop contribution. }

\item{We find the relation (\ref{aco}) truly remarkable.
Since the scale of inflation $ M $ is fixed, presumably by the
underlying microscopic (GUT) theory, the scaling of $ \lambda $
with the inverse of the number of e-folds strongly suggests a
\emph{renormalization group explanation of the effective field
theory} because the number of e-folds is associated with the
logarithm of the scale $ N=\ln a $. A renormalization group
improved scale dependent quartic coupling \cite{weipes} behaves  as
$ \lambda(K) \propto 1/\ln K $ with $ K $ the scale at which the
theory is studied. Since in an expanding cosmology the physical
scale grows with the scale factor it is natural to expect that a
renormalization group resummation yields the
renormalized coupling scaling as
$$
\lambda \sim 1/\ln a  \sim 1/N \; .
$$
}
\end{itemize}

There are several aspects of slow roll inflation, which when
considered together, lead us to conjecture that \emph{the
effective field theory of inflation is an almost critical theory
near but not at a trivial fixed point of the renormalization
group.} These aspects are the following:

\begin{itemize}
\item{The fluctuations of the inflaton are almost \emph{massless},
this is the statement that the slow roll parameter
$$
\eta_V = M^2_{Pl} \; \frac{V''(\phi) }{V(\phi)} \simeq 3 \;
\frac{V''(\phi)}{H^2} \ll 1 \; .
$$
The slow roll relation (\ref{etav}) states that the dimensionless
ratio of the inflaton mass and the Hubble scale is $\sim 1/N\sim 1/\ln a $.}

\item{The higher order couplings are suppressed by further
 powers of $ 1/N \sim 1/\ln a $ [see eq.(\ref{Gn})].
In the language of critical
 phenomena, the mass is a relevant operator in the infrared,
the quartic coupling $ \lambda $ is marginal,
 and higher order couplings are irrelevant. }

\end{itemize}

These ingredients taken together strongly suggest that for large
$ N\sim \ln a $, the effective field theory is reaching a
\emph{trivial gaussian infrared fixed point}\cite{clar}. The evidence for
this is manifest in that:

i) the power spectrum of scalar
fluctuations is \emph{nearly scale invariant} (a consequence of
$\eta_V \ll 1$),

ii)  the coupling constants
 vanish in the asymptotic limit $N \rightarrow \infty$. \emph{If}
 the number of e-folds were infinite, the theory would be
 sitting at the trivial fixed point: a massless free field
 theory.

\medskip

The physical situation in inflationary cosmology
is \emph{not} to be sitting \emph{exactly} at the fixed point,
 inflation \emph{must} end, and  is to be followed
by a radiation dominated standard big bang cosmology. Therefore,
 we conclude that \emph{during the stage of slow roll inflation, the
 theory is hovering near a trivial gaussian infrared fixed point}
but the presence of
a small relevant operator, namely the inflaton mass which
eventually becomes large at the end of slow roll, drives the
theory away from criticality.

\medskip

Our investigations\cite{clar} reveal that it is \emph{not} the
ultraviolet behavior of the  renormalization group that is
responsible for the near criticality of the effective field theory,
but rather the infrared, superhorizon physics. That this is the
case can be gleaned in eq. (\ref{aco}): the coefficient
$(M/M_{Pl})^4$ in front of the term $1/\ln a $ \emph{cannot} be
obtained from the usual Minkowski-space renormalization group solution for 
the running coupling. Furthermore, the true trivial fixed point is obtained in
the infrared limit when the scale factor $a\rightarrow \infty$, namely
infinitely long physical wavelengths.

\bigskip

If this conjecture\cite{clar} proves correct, it will have a major
fundamental appeal as a description of inflation because field
theories near a fixed point feature \emph{universal behavior}
independent of the underlying microscopic degrees of freedom. In
fact, if the effective field theory of
inflation is indeed near (not exactly) an infrared gaussian
fixed point, its predictions would be nearly
universal, and single field slow roll inflation describes a large
\emph{universality class} that features the same robust
predictions. Such is the case in critical phenomena
described by field theories where widely
different systems feature the same behavior near a critical point.

\section{Quantum corrections to the equation of motion for the
inflaton and its effective potential.}\label{eftsr}

We consider single field inflationary models described by a
general self-interacting scalar field theory in a spatially flat
Friedmann-Robertson-Walker cosmological space time with scale
factor $a(t)$. In comoving coordinates the action  is given by
\begin{equation}\label{action}
S= \int d^3x \; dt \;  a^3(t) \Bigg[ \frac{1}{2} \;
{\dot{\phi}^2}-\frac{(\nabla \phi)^2}{2a^2}-V(\phi) \Bigg] \;.
\end{equation}
We consider a \emph{generic} potential $V(\phi)$, the only
requirement is that its \emph{derivatives} be small in order to
justify the slow roll expansion\cite{hu,barrow}.
In order to study the corrections from the quantum fluctuations we
separate the classical homogeneous expectation value of the scalar
field from the quantum fluctuations by writing
\be\label{tad}
\phi(\vec{x},t)= \Phi_0(t)+\varphi(\vec{x},t)\;,
\ee
\noindent with
\be\label{exp}
\Phi_0(t)=\langle \phi(\vx,t) \rangle~~;~~ \langle
\varphi(\vx,t)\rangle =0 \;,
\ee
where the expectation value is in the non-equilibrium quantum
state. Expanding the Lagrangian density and integrating by parts, the
action becomes
\be\label{Split}
S= \int d^3x \; dt \;  a^3(t)\,
\mathcal{L}[\Phi_0(t),\varphi(\vx,t)]\;,
\ee
\noindent with
\bea\label{lagraQ}
&&\mathcal{L}[\Phi_0(t),\varphi(\vx,t)]  =
\frac{1}{2} \; {\dot{\Phi}^2_0}-V(\Phi_0)+\frac{1}{2} \;
{\dot{\varphi}^2}-\frac{(\nabla \varphi)^2}{2 \, a^2} -\frac{1}{2}\;
V^{''}(\Phi_0)\; \varphi^2 \cr \cr &&- \varphi\;
\left[\ddot{\Phi}_0+3 \, H \,\dot{\Phi}_0+V^{'}(\Phi_0)\right] -
\frac{1}{6}\; V^{'''}(\Phi_0)\; \varphi^3 - \frac{1}{24}\;
V^{(IV)}(\Phi_0)\; \varphi^4+ \textmd{higher orders in}\,
\varphi \; .
\eea
We obtain the equation of motion for the homogeneous
expectation value of the inflaton field  by implementing the tadpole
method (see \cite{pardec} and references
therein). This method consists in requiring the condition $\langle
\varphi(\vx,t)\rangle =0 $ consistently in a perturbative expansion
by treating the \emph{linear}, cubic, quartic (and higher order)
terms in the Lagrangian density eq.(\ref{lagraQ}) as
\emph{perturbations}\cite{pardec}.

Our approach relies on two distinct and fundamentally different
expansions: i) the effective field theory (EFT) expansion and  ii)
the slow-roll expansion.

Quantum corrections to the equations of motion for the inflaton and
for the fluctuations are 
obtained by treating the second line in eq.(\ref{lagraQ}),
namely, the \emph{linear} and the non-linear terms in $\varphi$, in
perturbation theory.

The generating functional of non-equilibrium real time correlation
functions requires a path integral along a complex contour in time:
the forward branch corresponds to time evolution forward $(+)$ and
backward $(-)$ in time as befits the time evolution of a density
matrix. Fields along these branches are labeled $\varphi^+$ and
$\varphi^-$, respectively (see
refs.\cite{pardec} and references therein).
The tadpole conditions 
\be\label{tads} 
\langle \varphi^\pm(\vx,t)
\rangle =0 \; , \ee \noindent both lead to the (same) equation of
motion for the expectation value $\Phi_0(t)$ by considering the
\emph{linear, cubic} and higher order terms in the Lagrangian
density as interaction vertices. To one loop order we find
\be\label{1lupeqn} 
\ddot{\Phi}_0(t)+3 \, H \; \dot{\Phi}_0(t)+V'(\Phi_0)+g(\Phi_0) \; \langle
[\varphi^+(\vx,t)]^2\rangle =0 \;. 
\ee 
The first three terms in eq.(\ref{1lupeqn}) are the familiar ones for the 
classical equation of motion of the inflaton.

The last term is the one-loop correction to the equations of motion
of purely quantum mechanical origin. Another derivation of this
quantum correction can be found in\cite{cosmo,ramsey}.
The fact that the tadpole method, which in this case results in a one-loop
correction, leads to a covariantly conserved and fully renormalized energy
momentum tensor has been established in the most general case in
refs.\cite{cosmo,erice,mottola}.

The coupling $g$, effective `mass term' $M^2$ and the quartic coupling
are defined by
\be \label{lambda}
M^2 \equiv M^2(\Phi_0)  =   V''(\Phi_0) = 3 \; H_0^2 \;
\eta_V + \mathcal{O}(\frac1{N}) \; , \; 
g\equiv g(\Phi_0)  =  \frac{1}{2} \;  V^{'''}(\Phi_0) \; , \; 
 \lambda \equiv \lambda (\Phi_0)  =  \frac{1}{6} \;
 V^{(IV)}(\Phi_0)\; . 
\ee
The $\langle(\cdots)\rangle$ is computed in the free
field (Gaussian) theory of the fluctuations $\varphi$ with an
effective `mass term' $M^2$, the
quantum state will be specified below. Furthermore, it is
straightforward to see that $\langle [\varphi^+(\vx,t)]^2\rangle =
\langle [\varphi^-(\vx,t)]^2\rangle=\langle
[\varphi(\vx,t)]^2\rangle$. In terms of the spatial Fourier
transform of the fluctuation field $\varphi(\vx,t)$, the one-loop
contribution can be written as 
\be\label{lupPS} 
\langle [\varphi(\vx,t)]^2\rangle = \int \frac{d^3 k}{(2\pi)^3} \; \langle
|\varphi_{\vk}(t)|^2 \rangle = \int_0^{\infty} \frac{dk}{k} \;
\mathcal{P}_{\varphi}(k,t)\,, 
\ee 
\noindent where $\varphi_{\vk}(t)$
is the spatial Fourier transform of the fluctuation field
$\varphi(\vx,t)$ and we have introduced the power spectrum of the
fluctuation 
\be\label{PS} \mathcal{P}_{\varphi}(k,t) = \frac{k^3}{2
\, \pi^2} \; \langle |\varphi_{\vk}(t)|^2 \rangle \,. 
\ee
During slow roll inflation the scale factor is quasi de Sitter and
to lowest order in slow roll it is given by : 
\be\label{quasiDS}
C(\eta)=-\frac{1}{H_0 \; \eta} \; \frac{1}{1-\epsilon_V}=
-\frac{1}{H_0 \; \eta} (1+\epsilon_V) + \mathcal{O}(\epsilon_V^2) \,. 
\ee 
The spatial Fourier transform of the rescaled free field Heisenberg
operators $\chi(\vx,\eta) \equiv C(\eta) \varphi(\vx,t) $ obey the equation 
\be\label{heiseqn}
\chi^{''}_{\vk}(\eta)+ \left[k^2 + M^2 \;  C^2(\eta)-
\frac{C^{''}(\eta)}{C(\eta)} \right]\chi_{\vk}(\eta)=0 \,.
\ee
Using the slow roll expressions eqs.(\ref{lambda}) and (\ref{quasiDS}),
it becomes
\be\label{heiseqn2}
\chi^{''}_{\vk}(\eta)+ \left[k^2
-\frac{\nu^2-\frac{1}{4}}{\eta^2} \right]\chi_{\vk}(\eta)=0 
\ee
\noindent where the index $ \nu $ and the quantity $ \Delta $ are 
given by 
\be\label{nu} \nu = \frac{3}{2} + \epsilon_V-\eta_V
+\mathcal{O}(\frac1{N^2}) \quad , \quad
\Delta= \frac{3}{2}-\nu = \eta_V-\epsilon_V +\mathcal{O}(\frac1{N^2}) \; .
\ee
The scale invariant case $ \nu = \frac{3}{2} $ corresponds to
massless inflaton fluctuations in the de Sitter background. 
$ \Delta $ measures the departure from scale invariance. 
In terms of the spectral index
of the scalar adiabatic perturbations $ n_s $ and the ratio $ r $ of
tensor to  scalar perturbations, $ \Delta $ takes the form, \be
\Delta=\frac12 \left( n_s - 1 \right) + \frac{r}8 \; . \ee The free
Heisenberg field operators $\chi_{\vk}(\eta)$ are written in terms
of annihilation and creation operators that act on Fock states  as
\be\label{ope} \chi_{\vk}(\eta) = a_{\vk} \; S_{\nu}(k,\eta)+
a^{\dagger}_{-\vk} \; S^{*}_{\nu}(k,\eta) \ee \noindent where the
mode functions $S_{\nu}(k,\eta)$ are solutions of the eqs.
(\ref{heiseqn2}). For Bunch-Davis boundary conditions we have
\be\label{BDS} 
S_{\nu}(k,\eta) = \frac{1}{2}
\; \sqrt{-\pi\eta} \; e^{i\frac{\pi}{2}(\nu+\frac{1}{2})} \;
H^{(1)}_\nu(-k\eta)\, , 
\ee 
this defines the Bunch-Davis vacuum $ a_{\vk} |0>_{BD} =0 $.

There is no unique choice of an initial state, and a recent body of work
has began to address this issue (see ref.\cite{mottola}
for a discussion and further references). A full study of the
\emph{quantum loop} corrections with different initial
states must first elucidate the behavior of the propagators for the
fluctuations in such states. Here we focus on the standard
choice in the literature\cite{hu} which allows us to include the
quantum corrections into the standard results in the literature. A study
of quantum loop corrections with different initial states is an important
aspect by itself which we postpone to later work.

The index $\nu$ in the mode functions eq.(\ref{BDS}) depends on the
expectation value of the scalar field,  via the slow roll variables,
hence it slowly varies in time. Therefore, it is  consistent to treat
this time  dependence of  $\nu$ as an \emph{adiabatic approximation}.
This is well known and standard in the slow roll
expansion\cite{hu}.
Indeed, there are corrections to the mode functions which are higher order
in slow roll.
However, these mode functions enter in the propagators in loop corrections,
therefore they yield higher order contributions in slow roll
and we discard them consistently to lowest order in slow roll.

With this choice and to lowest order in slow roll, the power
spectrum eq.(\ref{PS}) is given by \be\label{PSSR}
\mathcal{P}_{\varphi}(k,t) = \frac{H^2}{8 \, \pi} \; (-k\eta)^3 \;
|H^{(1)}_\nu(- k \eta)|^2 \,. \ee For large momenta $|k\eta| \gg 1$
the mode functions behave just like free field modes in Minkowski
space-time, namely \be S_{\nu}(k,\eta) \buildrel{|k\eta| \gg
1}\over= \frac{1}{\sqrt{2k}} \; e^{-ik\eta} \ee \noindent Therefore,
the quantum correction to the equation of motion for the inflaton
eqs.(\ref{1lupeqn}) and (\ref{lupPS}) determined by the momentum
integral of $ \mathcal{P}_{\varphi}(k,t) $ features both quadratic
and logarithmic divergences. Since the field theory inflationary
dynamics is an \emph{effective field theory} valid below a comoving
cutoff $\Lambda$ of the order of the Planck scale, the one loop
correction (\ref{lupPS}) becomes \be\label{PSint} \int^{{\Lambda}}_0
\frac{dk}{k} \,\mathcal{P}_{\varphi}(k,t) = \frac{H^2}{8 \, \pi}
\int^{\Lambda_p}_0 \frac{dz}{z} \; z^3\,
\left|H^{(1)}_\nu(z)\right|^2 \,, \ee \noindent where $
\Lambda_p(\eta)$ is the ratio of the cutoff in physical coordinates
to the scale of inflation, namely \be\label{physcut} \Lambda_p(\eta)
\equiv\frac{\Lambda}{H \; C(\eta)}=-\Lambda\,  \eta\;. \ee The
integration variable $ z=-k \, \eta $ has a simple interpretation at
leading order in slow roll \be\label{zSR} z \equiv -k \, \eta =
\frac{k}{H_0 \, C(\eta)}= \frac{k_p(\eta)}{H_0} \,, \ee \noindent
where $k_p(\eta)=k/C(\eta)$ is the wavevector in physical
coordinates. If the spectrum of scalar fluctuations were strictly
scale invariant, (namely for massless inflaton fluctuations in de
Sitter space-time), then the index would be $\nu=3/2$ and  the
integrand in (\ref{PSint}) given by \be\label{integ} z^3 \,
\left|H^{(1)}_{\frac{3}{2}}(z)\right|^2 =
\frac{2}{\pi}\left[1+z^2\right]\,. \ee In this strictly scale
invariant case, the integral of the power spectrum also features an
\emph{infrared} logarithmic divergence. While the ultraviolet
divergences are absorbed by the renormalization counterterms in the
effective field theory, this is not possible for the
infrared divergence. Obviously, the origin of this infrared behavior
is the {\bf exact} scale invariance of superhorizon fluctuations.
However, during slow roll inflation there are small corrections to
scale invariance, in particular the index $\nu$ is slightly different 
from $3/2$ and this slight departure introduces a natural infrared
regularization. In ref.\cite{pardec} we have
introduced an expansion in the parameter $\Delta = 3/2-\nu=
\eta_V-\epsilon_V+\mathcal{O}(\epsilon^2_V,\eta^2_V,\epsilon_V\eta_V)$
which is small during slow roll and we showed in \cite{qua} 
that the infrared divergences featured by
the quantum correction manifest as \emph{simple poles} in $\Delta$.

The quantum correction to the equation of motion for the inflaton by 
isolating the pole in $\Delta$ as well as the leading logarithmic divergences  
were computed in ref.\cite{qua} with the result
\be\label{QC}
\frac12 \langle[\varphi(\vx,t)]^2\rangle = \left(\frac{H_0}{4 \, \pi}\right)^2
\left[ {\Lambda_p}^2 + \ln \Lambda_p^2 +\frac1{\Delta}
 + 2 \, \gamma - 4 + \mathcal{O}(\Delta) \right]\,,
\ee
\noindent
where $\gamma$ is the Euler-Mascheroni constant.  While the
quadratic and logarithmic \emph{ultraviolet} divergences are
regularization scheme dependent, the pole in $\Delta$ arises from
the  infrared behavior and is independent of the regularization
scheme. In particular this pole coincides with that found
in the expression for $<\phi^2(\vx,t)>$ in ref.\cite{fordbunch}. The
\emph{ultraviolet divergences}, in whichever renormalization scheme,
 require that the effective field theory be
defined to contain \emph{renormalization counterterms} in the bare
effective lagrangian, so that these counterterms will systematically
cancel the divergences encountered in the calculation of quantum
corrections in the (EFT) and slow roll approximations.

\subsection{Renormalized effective field theory: renormalization counterterms}

The renormalized effective field theory is obtained by writing the
potential $V[\phi]$ in the Lagrangian density eq.(\ref{action}) in
the following form \be\label{count}
V(\phi)=V_R(\phi)+\delta\,V_R(\phi,\Lambda)\;, \ee \noindent where
$V_R(\phi)$ is the renormalized \emph{classical} inflaton potential
and $\delta\,V_R(\phi,\Lambda)$ includes the renormalization
counterterms which are found systematically in a slow roll expansion
by canceling the ultraviolet divergences. In
this manner, the equations of motion and correlation functions in
this effective field theory \emph{are cutoff independent}. 
We find from eqs.(\ref{1lupeqn}) and (\ref{QC}) 
\be\label{equalup}
\ddot{\Phi}_0(t)+3\,H\,\dot{\Phi}_0(t)+V'(\Phi_0)+ V^{'''}(\Phi_0)
\left(\frac{H_0}{4 \, \pi}\right)^2 \left[\Lambda_p^2+
\ln\Lambda_p^2 +\frac{1}{\Delta}+2 \, \gamma - 4 +
\mathcal{O}(\Delta)\right]=0 \; . 
\ee
\noindent From this equation it becomes clear that the one-loop ultraviolet
divergences can be canceled by choosing appropriate counterterms\cite{qua}
leading to the final form of the
renormalized inflaton equation of motion to leading order in the
slow roll expansion 
\be\label{fineq}
\ddot{\Phi}_0(t)+3\,H_0\,\dot{\Phi}_0(t)+V^{'}_R(\Phi_0)+
\left(\frac{H_0}{4 \, \pi}\right)^2
\frac{V^{'''}_R(\Phi_0)}{\Delta}=0 \;. 
\ee 
Although the quantum correction is
of order $ V^{'''}_R(\Phi_0) $, (second order in slow
roll),  the strong infrared divergence arising from the quasi
scale invariance
 of inflationary fluctuations brings about a denominator
which is of first order in slow roll. Hence the lowest order
quantum correction in the slow roll expansion,
is actually  of the same order as $ V^{'}_R(\Phi_0) $.
To highlight this observation, it proves convenient to write
eq.(\ref{fineq}) in terms of the EFT and slow roll parameters,
\be\label{fineqsr}
\ddot{\Phi}_0(t)+3\,H_0\,\dot{\Phi}_0(t)+
V^{'}_R(\Phi_0)\left[1+\left(\frac{H_0}{2\pi \, M_{Pl}}\right)^2
\frac{\xi_V}{2\,\epsilon_V\,\Delta}\right]=0 \;.
\ee
Since $\xi_V \sim \epsilon^2_V$ and $\Delta \sim \epsilon_V$ the leading quantum
corrections are of zeroth order in slow roll. This is a consequence
of the infrared enhancement resulting from the nearly scale
invariance of the power spectrum of scalar fluctuations. The quantum
correction is  suppressed by an EFT factor $H^2/M^2_{Pl} \ll 1$.

Restoring the dependence of $\Delta$ on $\Phi_0$ through the
definitions (\ref{etav}) and (\ref{nu}) we
find the equation of motion for the inflaton field
to leading order in slow roll and in $ (H/M_{Pl})^2 $,
\be\label{eqnslor}\ddot{\Phi}_0(t)+3\,H\,\dot{\Phi}_0(t)+
V^{'}_R(\Phi_0)+\frac{1}{24 \, (\pi \;  M_{Pl})^2} \;
\frac{V_R^3(\Phi_0)\,
V^{'''}_R(\Phi_0)}{2 \, V_R(\Phi_0)V^{''}_R(\Phi_0)-V^{'\,2}_R(\Phi_0)
}=0 \;.
\ee

\subsection{Quantum corrections to the Friedmann equation:
the effective potential}\label{friedeq}

The zero temperature effective potential in Minkowski space-time is often
used to describe the scalar field dynamics during inflation
\cite{hu,riottorev}. However, as we see below  the resulting effective potential
[see eq.(\ref{Veff})]  is remarkably different from the Minkowski one
[see Appendix A]. The focus of this Section is to derive the effective potential for
slow-roll inflation.

Since the fluctuations of the inflaton field are quantized, the
interpretation of the `scalar condensate' $\Phi_0$ is that of the
expectation value of the full quantum field $\phi$ in a homogeneous
coherent quantum state. Consistently with this, the Friedmann
equation must necessarily be understood in terms of the
\emph{expectation} value of the field energy momentum tensor, namely
\be\label{FRW2} 
H^2= \frac{1}{3 \, M^2_{Pl}}\left\langle \frac{1}{2}
\; \dot{\phi}^2+\frac{1}{2} \; \left(\frac{\nabla
\phi}{a(t)}\right)^2+V[\phi] \right\rangle  \;. 
\ee 
Separating the homogeneous condensate from the fluctuations as in eq. (\ref{tad})
and imposing the tadpole equation (\ref{exp}), the Friedmann equation becomes
\be\label{FRexp} 
H^2= \frac{1}{3 \, M^2_{Pl}}\left[ \frac{1}{2} \;
{\dot{\Phi_0}}^2 + V_R(\Phi_0)+\delta V_R(\Phi_0)\right]+ \frac{1}{3
\, M^2_{Pl}}\left\langle \frac{1}{2} \; \dot{\varphi}^2+\frac{1}{2}
\; \left(\frac{\nabla \varphi}{a(t)}\right)^2+\frac{1}{2} \;
V^{''}(\Phi_0)\; \varphi^2 +\cdots\right\rangle 
\ee
The dots inside the angular brackets correspond to terms with higher
derivatives of the potential which are  smaller in the slow roll
expansion. The quadratic term $\langle \varphi^2 \rangle$ 
to leading order in slow roll is given by eq.(\ref{QC}). 
Calculating the expectation value in eq.(\ref{FRexp}) in
free field theory  corresponds to obtaining the corrections to the
energy momentum tensor by integrating the fluctuations \emph{up to
one loop}\cite{qua}. 
The first two terms of the expectation value in  eq.(\ref{FRexp})
\emph{do not} feature infrared divergences for $\nu=3/2$ because of
the two extra powers of the loop momentum in the integral. These
contributions are given by 
\bea\label{kinterm} 
&&\left\langle
\frac{1}{2} \; \dot{\varphi}^2 \right\rangle = \frac{H^4_0}{16
\,\pi} \; \int^{\Lambda_p}_{0} \frac{dz}{z} \;  z^2 \;
\left|\frac{d}{dz}\left[z^{\frac{3}{2}} H^{(1)}_{\nu}(z) \right]
\right|^2 = \frac{H^4_0 \; \Lambda^4_p}{32 \,\pi^2}+
\mathcal{O}( H^4_0 \Delta)\;,\\
\label{grad} &&\left\langle\frac{1}{2}\left(\frac{\nabla
\varphi}{a(t)}\right)^2 \right \rangle = \frac{H^4_0}{16 \,\pi} \;
\int^{\Lambda_p}_{0} \frac{dz}{z} \; z^{5} \;  \left|
H^{(1)}_{\nu}(z)  \right|^2 = \frac{H^4_0 \; \Lambda^4_p}{32 \,
\pi^2}+ \frac{H^4 \; \Lambda^2_p}{16 \, \pi^2} +\mathcal{O}( H^4_0
\Delta)\;. 
\eea 
The counterterms cancel the ultraviolet divergences arising from the third term 
in the angular brackets in eq. (\ref{FRexp})\cite{qua}. Finally,
the fully renormalized Friedmann equation to one loop and to lowest order
in the slow roll expansion is \cite{qua}
\be\label{FRren} H^2 =
\frac{1}{3 \, M^2_{Pl}}\left[ \frac{1}{2} \; {\dot{\Phi_0}}^2 +
V_R(\Phi_0) + \left(\frac{H_0}{4 \,
\pi}\right)^2\frac{V^{''}_R(\Phi_0)}{\Delta} +\textmd{higher orders
in slow roll}\right] \equiv H^2_0 + \delta H^2 \;, \ee \noindent
where $H_0$ is the  Hubble parameter in absence of quantum
fluctuations:
$$
 H^2_0 = \frac{V_R(\Phi_0)}{3 \, M^2_{Pl}} \left[1+\frac{\epsilon_V}{3}+
\mathcal{O}(\epsilon^2_V,\epsilon_V \; \eta_V) \right] \; .
$$
Using the lowest order slow roll relation eq. (\ref{lambda}), the
last term in eq.(\ref{FRren}) can be written as follows
\be\label{delH}
\frac{\delta H^2}{H^2_0} = \left(\frac{H_0}{4 \,
\pi\,M_{Pl}}\right)^2 \frac{\eta_V}{\Delta}\;.
\ee
This equation defines the back-reaction correction to the scale factor
arising from the quantum fluctuations of the inflaton.

Hence, while the ratio $\eta_V/\Delta$ is of order zero in slow roll,
the one loop correction to the Friedmann equation is of the order
$H^2_0/M^2_{Pl} \ll 1$ consistently with the EFT expansion.
The Friedmann equation suggests the identification of the effective potential
\bea\label{Veff}
&&V_{eff}(\Phi_0) = V_R(\Phi_0)+ \left(\frac{H_0}{4
\, \pi}\right)^2\frac{V^{''}_R(\Phi_0)}{\Delta} +\textmd{higher
orders in slow roll} = \\ \cr
&&= V_R(\Phi_0)\left[1+
\left(\frac{H_0}{4 \, \pi\,M_{Pl}}\right)^2\frac{\eta_V}{
\eta_V-\epsilon_V} + \textmd{higher orders in slow roll} \right]
 \; . \label{Vefsr}
\eea
We see that the equation of motion for the inflaton eq.(\ref{fineq})
takes the natural form
$$
\ddot{\Phi}_0(t)+3\,H_0\,\dot{\Phi}_0(t)+
\frac{\partial V_{eff}}{\partial\Phi_0}(\Phi_0)
= 0 \;.
$$
where the derivative of $ V_{eff} $ with respect to $ \Phi_0 $
is taken at fixed Hubble and slow roll parameters. That is,
$ H_0 $ and $ \Delta $ must be considered in the present context
as gravitational degrees of freedom and not as matter (inflaton)
degrees of freedom.

Eqs.(\ref{delH}) and (\ref{Veff}) make manifest the nature of the
effective field theory expansion in terms of the ratio
$ \left(H_0/M_{Pl}\right)^2 $. The coefficients of the powers of this
ratio are obtained in the slow roll expansion. To leading order,
these coefficients are of $ \mathcal{O}(N^0) $ because of the
infrared enhancement manifest in the poles in $\Delta$, a
consequence of the nearly scale invariant power spectrum of scalar
perturbations.

A noteworthy result is the rather different form of the effective
potential eq.(\ref{Veff}) as compared to the result in Minkowski
space time at zero temperature.
In the appendix we show explicitly that the same
definition of the effective potential as the expectation value of
$T_{00}$ in Minkowski space-time  at zero temperature 
is strikingly  different from eq.(\ref{Veff}) valid for slow roll inflation.

\section{Quantum Corrections to the Scalar and Tensor Power.
Scalar, Tensor, Fermion and Light Scalar Contributions.}

During slow-roll inflation many quantum fields may be coupled to the inflaton
(besides itself) and can contribute to the quantum corrections to the equations
of motion and to the inflaton effective potential. The scalar curvature
and tensor fluctuations are certainly there. We also consider light fermions and 
scalars coupled to the inflaton. 
We take the fermions to be Dirac fields with a generic Yukawa-type coupling
but it is straightforward to generalize to Weyl or Majorana fermions. 
The Lagrangian density is taken to be
\be
\mathcal{L} = \sqrt{-g} \; \Bigg\{ \frac12 \;  \dot{\varphi}^2 -
\left(\frac{\vec{\nabla}\varphi}{2\,a}\right)^2-V(\varphi)+\frac12 \;
\dot{\sigma}^2 - \left(\frac{\vec{\nabla}\sigma}{2\,a}\right)^2 -
\frac12 \;  m^2_\sigma \;  \sigma^2 -G(\varphi)  \;  \sigma^2+
\overline{\Psi}\Big[i\,\gamma^\mu \;  \mathcal{D}_\mu \Psi -m_f -
Y(\varphi)\Big]\Psi \Bigg\} \label{lagrangian}
\ee
\noindent where $G(\Phi)$ and $Y(\Phi)$ are generic interaction terms
between the inflaton and the scalar and fermionic fields.
For simplicity we
consider one bosonic and one fermionic Dirac field.
The $\gamma^\mu$ are the curved space-time Dirac matrices
and $ \mathcal{D}_\mu $ the fermionic covariant derivative\cite{anom,BD}.
We will consider that the light scalar field $\sigma$ has vanishing
expectation value at all times, therefore inflationary dynamics is
driven by one single scalar field, the inflaton $\phi$. 

We consider the contributions from the quadratic fluctuations
to the energy momentum tensor. There are \emph{four} distinct
contributions: i) scalar metric (density) perturbations, ii) tensor
(gravitational waves) perturbations, iii) fluctuations of the light
bosonic scalar field $\sigma$, iv) fluctuations of the light
fermionic field $\Psi$.

Fluctuations in the metric are studied as
usual \cite{hu,mukhanov}. Writing the metric as
$$
g_{\mu\nu}= g^0_{\mu\nu}+\delta^s g_{\mu\nu}+\delta^t g_{\mu\nu}
$$
\noindent where $g^0_{\mu\nu}$ is the spatially flat FRW background
metric eq.(\ref{metrica}), $\delta^{s,t} g_{\mu\nu}$ correspond
to the scalar and tensor perturbations respectively, and we neglect
vector perturbations. In longitudinal gauge
\be
\delta^{s} g_{00}  =  C^2(\eta) \; 2  \; \phi \quad , \quad
\delta^{s} g_{ij}  =  C^2(\eta) \;  2 \;  \psi \;  \delta_{ij}
\label{curvpot} \quad , \quad
\delta^{t} g_{ij}  =  -C^2(\eta) \;  h_{ij} 
\ee 
\noindent where $ h_{ij} $ is transverse and
traceless and we neglect vector modes since they are not
generated in single field inflation\cite{hu,mukhanov}.

Expanding up to quadratic order in the
scalar fields, fermionic fields and metric perturbations the part
of the Lagrangian density that is quadratic in these fields is
given by
$$
\mathcal{L}_Q =\mathcal{L}_s[\delta\varphi^{gi},\phi^{gi},\psi^{gi}]+
\mathcal{L}_t[h]+\mathcal{L}_\sigma[\sigma]+
\mathcal{L}_\Psi[\overline{\Psi},\Psi] \; ,
$$
\noindent where
\bea
&&\mathcal{L}_t[h] = \frac{M^2_{Pl}}{8} \; C^2(\eta) \;
\partial_\alpha h^j_i  \; \partial_\beta h^i_j \;  \eta^{\alpha \beta} \; ,
\cr \cr &&
\mathcal{L}_\sigma[\sigma]=
 C^4(\eta) \; \Bigg\{\frac{1}{2} \;
\left(\frac{\sigma'}{C}\right)^2-\frac{1}{2} \; \left(\frac{\nabla
\sigma}{ C}\right)^2 -\frac{1}{2} \; M^2_\sigma[\Phi_0]\;
\sigma^2\Bigg\}  \; , \cr \cr && 
\mathcal{L}_\Psi[\overline{\Psi},\Psi]= \overline{\Psi}\Big[i \;
\gamma^\mu  \; \mathcal{D}_\mu \Psi -M_\Psi[\Phi_0]\Big]\Psi \; ,
\nonumber 
\eea \noindent 
here the prime stands for derivatives
with respect to conformal time and the labels (gi) refer to gauge
invariant quantities\cite{mukhanov}. The explicit expression for
$\mathcal{L}[\delta\varphi^{gi},\phi^{gi},\psi^{gi}]$ is given in
eq. (10.68) in ref.\cite{mukhanov}. The effective masses for the
bosonic and fermionic fields are given by 
\be
M^2_\sigma[\Phi_0]  =  m^2_\sigma + G(\Phi_0) \label{sigmamass}
\quad , \quad 
M_\Psi[\Phi_0]  =  m_f+Y(\Phi_0) \; . 
\ee
We focus on the study of the quantum corrections to the Friedmann
equation, for the case in which both the scalar and fermionic
fields are light in the sense that during slow roll inflation, 
\be
M_\sigma[\Phi_0]\; , \; M_\Psi[\Phi_0] \ll H_0 \; , 
\ee 
\noindent at least during the cosmologically relevant stage corresponding to
the 60 or so e-folds before the end of inflation.

From the quadratic Lagrangian given above the  quadratic quantum
fluctuations to the energy momentum tensor can be extracted.

The effective potential is identified with $\langle T^0_0 \rangle$
in a spatially translational invariant state in which the
expectation value of the inflaton field is $\Phi_0$. During slow
roll inflation the expectation value $\Phi_0$ evolves very slowly
in time, the slow roll approximation is indeed an adiabatic
approximation, which justifies treating $\Phi_0$ as a constant in
order to obtain the effective potential. The time variation of
$\Phi_0$ only contributes to higher order corrections in
slow-roll. The energy momentum tensor is computed in the
FRW inflationary background determined by the \emph{classical}
inflationary potential $V(\Phi_0)$, and the slow roll parameters
are also explicit functions of $\Phi_0$. Therefore the energy
momentum tensor depends \emph{implicitly} on $\Phi_0$ through the
background and \emph{explicitly} on the masses for the light
bosonic and fermionic fields given above.

We can write the effective potential as
\be 
V_{eff}(\Phi_0) = V(\Phi_0)+ \delta V(\Phi_0) \; , \label{Veff2} 
\ee 
\noindent where 
\be
\delta V(\Phi_0)= \langle T^{0}_{0}[\Phi_0] \rangle_s + \langle
T^{0}_{0}[\Phi_0] \rangle_t + \langle T^{0}_{0}[\Phi_0]
\rangle_\sigma +\langle T^{0}_{0}[\Phi_0]
\rangle_\Psi\label{dVeff} 
\ee 
\noindent $(s,t,\sigma,\Psi)$
correspond to the energy momentum tensors of the quadratic
fluctuations of the scalar metric, tensor (gravitational waves),
light boson field $\sigma$ and light fermion field $\Psi$
fluctuations respectively. Since these are the expectation values
of a quadratic energy momentum tensor, $\delta V(\Phi_0)$
corresponds to the \emph{one loop correction} to the effective
potential.

\subsection{Light scalar fields}

During slow roll inflation the effective mass of the $\sigma$
field is given by eq. (\ref{sigmamass}), just as for the inflaton
fluctuation in sec. 2. It is convenient to introduce a parameter
$\eta_\sigma$ defined to be 
\be\label{etasig} \eta_\sigma =
\frac{M^2_\sigma[\Phi_0]}{3 \;  H^2_0 } \; .
\ee 
Hence, the sigma field contributions to the inflaton equations of motion
and inflaton effective potential can be obtained from sec. 2 just replacing
the slow roll parameter $ \eta_V $ by $ \eta_\sigma $. In particular,
infrared divergences are now regulated by the parameter $ \Delta_\sigma \equiv
\eta_\sigma - \epsilon_V $. 

To leading order in the slow roll expansion and in
$\eta_\sigma \ll 1$, the infrared contribution is given by\cite{anom},
$$
\frac{M^2_\sigma[\Phi_0]}{2} \; \big \langle \sigma^2(\vec{x},t) \big
\rangle =
 \frac{3\,H^4_0}{(4 \; \pi)^2} \; \frac{\eta_\sigma}{ \eta_\sigma-\epsilon_V}+
\mathrm{subleading  ~in ~ slow ~ roll}.
$$
The fully renormalized contribution  from from the sigma field to $ T^0_0 $
to leading order in slow roll takes the form \cite{anom},
 \be
\langle T^0_0 \rangle_\sigma =
 \frac{3\,H^4_0}{(4 \; \pi)^2}\frac{\eta_\sigma}{ \Delta_\sigma}+
\frac{1}{2}\,\Bigg\langle\dot{\sigma}^2+ \left(\frac{\nabla
\sigma}{C(\eta)}\right)^2 \Bigg\rangle_{ren} \label{T00siglead}
\ee
In calculating here the second term we can set to zero the slow
roll parameters $\epsilon_V, \eta_V$ as well as the mass of the
light scalar, namely $\eta_\sigma=0$. Hence, to leading order,
the second term is identified with the $00$ component of the
renormalized energy momentum tensor for a free massless minimally
coupled scalar field in exact de Sitter space time. Therefore 
we can extract this term from refs.\cite{fordbunch,BD},
\be\label{TmunudS}
\langle T_{\mu \nu}\rangle_{ren} = \frac{g_{\mu
\nu}}{(4 \, \pi)^2}\Bigg\{m^2_\sigma \; H^2_0
\left(1-\frac{m^2_\sigma}{2 \, H^2_0}\right)\left[-\psi\left(\frac{3}{2}+
\nu\right)-\psi\left(\frac{3}{2}-\nu\right)+
\ln\frac{m^2_\sigma}{H^2_0} \right]+
\frac{2}{3} \; m^2_\sigma  \; H^2_0-\frac{29}{30} \; H^4_0 \Bigg\} \; , 
\ee
where $ \nu \equiv \sqrt{\frac{9}{4}-\frac{m^2_\sigma}{H^2_0}} $ and 
$ \psi(z) $ 
stands for the digamma function. This expression corrects a factor of two in
ref.\cite{BD}. In eq. (6.177) in \cite{BD} the
D'Alambertian acting on $G^{1}(x,x')$ was neglected. However, in
computing this term, the D'Alambertian must be calculated
\emph{before} taking the coincidence limit. Using the equation of
motion yields the extra factor 2 and the expression eq.(\ref{TmunudS}).

The pole at $ \nu=3/2 $ manifest in eq.(\ref{TmunudS})
coincides with the pole in eq.(\ref{T00siglead}) using that
$ m^2_\sigma = 3 \; H^2 \;  \eta_\sigma $ [eq.(\ref{etasig})]. 
This pole originates in the term $m_\sigma^2 \;
<\sigma^2>$, which features an infrared divergence in the 
limit $\nu_\sigma=3/2$. All the terms with space-time derivatives are infrared
finite in this limit. Therefore, we can extract from eq.(\ref{TmunudS}) 
the renormalized expectation value in the limit $H_0 \gg m_\sigma$, 
\be 
\langle T^0_0 \rangle_\sigma =
 \frac{H^4_0}{(4 \; \pi)^2}\left[\frac{3\,\eta_\sigma}{
 \eta_\sigma-\epsilon_V}-\frac{29}{30}+
\mathcal{O}(\epsilon_V,\eta_\sigma,\eta_V)\right]\label{Tsiglea}
\ee
The second term is completely determined by the {\bf trace 
anomaly} of the minimally coupled scalar fields\cite{fordbunch,BD,ddhf}.

\subsection{Quantum Corrections to the Inflaton potential from
the Scalar metric perturbations}

The gauge invariant energy momentum tensor for quadratic scalar
metric fluctuations has been obtained in ref.\cite{abramo}. 
In longitudinal gauge and in cosmic time it is given  by
\bea
\langle T^0_0 \rangle_s = && M^2_{Pl} \Bigg[12 \; H_0 \;  \langle \phi
\dot{\phi} \rangle - 3 \;  \langle (\dot{\phi})^2 \rangle +
\frac{9}{C^2(\eta)} \; \langle (\nabla \phi)^2\rangle \Bigg] \nonumber \\
& & + \frac{1}{2} \; \langle (\dot{\delta \varphi})^2\rangle + \frac{\langle
(\nabla \delta \varphi)^2\rangle}{2\,C^2(\eta)} + \frac{V''(\Phi_0)}{2} \;
\langle(\delta \varphi)^2\rangle + 2 \;  V'(\Phi_0) \;
\langle \phi \,\delta \varphi
\rangle \label{T00s}
\eea
\noindent where the dots stand for derivatives with respect to cosmic time.
In longitudinal gauge, the equations of motion in cosmic time for
the Fourier modes are\cite{hu,mukhanov} 
\bea \label{phieq}
&&\ddot{\phi}_{\vec k}+ \left(H_0-2 \;
\frac{\ddot{\Phi}_0}{\dot{\Phi}_0}\right)\dot{\phi}_{\vec k}+
\left[2 \; \left(\dot{H}_0-2 \; H_0 \;
\frac{\ddot{\Phi}_0}{\dot{\Phi}_0}\right)+
\frac{k^2}{C^2(\eta)}\right]{\phi}_{\vec k}=0 \cr \cr &&
\label{delfieqn} \ddot{\delta \varphi}_{\vec k}+3 \; H \;
\dot{\delta \varphi}_{\vec
k}+\left[V''[\Phi_0]+\frac{k^2}{C^2(\eta)} \right]\delta
\varphi_{\vec k}+2 \;  V'[\Phi_0] \; \phi_{\vec k}- 4 \;
\dot{\Phi}_0 \; \dot{\phi}_{\vec k}=0  \; , 
\eea 
\noindent with the constraint equation 
\be \label{constraint} \dot{\phi}_{\vec
k}+H_0 \; \phi_{\vec k}= \frac{1}{2M_{Pl}} \; \delta \varphi_{\vec
k} \; \dot{\Phi}_0  \; . 
\ee 
We split the contributions to the energy momentum tensor as those
from superhorizon modes, which yield the infrared
enhancement, and the subhorizon modes for which we can set all
slow roll parameters to zero. Just as discussed above for the case
of the $\sigma$ field, since spatio-temporal derivatives bring
higher powers of the momenta, we can neglect all derivative terms
for the contribution from the superhorizon modes and set\cite{abramo} 
\be \langle
T^0_0 \rangle_{IR} \approx \frac{1}{2} \;  V''[\Phi_0] \; \langle
\left(\delta \varphi (\vec{x},t)\right)^2 \rangle + 2 \;
V'[\Phi_0] \;  \langle \phi(\vec{x},t)\,\delta \varphi(\vec{x},t)
\rangle  \; . 
\ee 
The analysis of the solution of eq.(\ref{phieq})
for superhorizon wavelengths in ref. \cite{mukhanov} shows that in
exact de Sitter space time $\phi_{\vec k} \sim \mathrm{constant}$,
hence it follows that during quasi-de Sitter slow roll inflation
for superhorizon modes 
\be \label{fipun} \dot{\phi}_{\vec k} \sim
(\mathrm{slow~roll}) \times H_0 \; \phi_{\vec k} 
\ee Therefore,
for superhorizon modes, the  constraint equation
(\ref{constraint}) yields \be \label{rela} \phi_{\vec k} = -\,
\frac{V'(\Phi_0)}{2 \; V(\Phi_0)}  \;  \delta \varphi_{\vec k} +
{\rm higher ~ orders ~ in ~ slow ~ roll} \; . \ee Inserting this
relation in eq.(\ref{delfieqn}) and consistently neglecting the
term $\dot{\phi}_{\vec k}$ according to eq.(\ref{fipun}), we find
the following equation of motion for the gauge invariant scalar
field fluctuation in longitudinal gauge 
\be \label{delfieqn2}
\ddot{\delta \varphi}_{\vec k}+3 \; H_0  \; \dot{\delta
\varphi}_{\vec k}+\left[\frac{k^2}{C^2(\eta)}+3 \;  H^2_0 \;
\,\eta_\delta \right]\delta \varphi_{\vec k}=0  \; , 
\ee \noindent
where we have used the definition of the slow roll parameters
$\epsilon_V; \; \eta_V$ given in eq.(\ref{etav}), and introduced
$ \eta_\delta \equiv \eta_V-2 \; \epsilon_V $.
This is the equation of motion for a minimally coupled scalar
field with mass squared $3 \;  H^2_0  \; \eta_\delta$ and we can
use the results obtained in the case of the scalar field $\sigma$. 

Repeating the analysis presented
in the case of the scalar field $\sigma$ above, we finally find\cite{anom}
\be 
\langle T^0_0 \rangle_{IR} = \frac{3 \; H^4_0}{(4 \; \pi)^2}
\; \frac{\eta_V-4\,\epsilon_V}{\eta_V-3 \, \epsilon_V} +
\mathrm{subleading~in~slow~roll} 
\ee 
It can be shown that the contribution from subhorizon modes to 
$ \langle T^0_{0s} \rangle$ is given by \cite{anom}
\be \langle T^0_{0s}\rangle_{UV} \simeq
\frac{1}{2}\langle (\dot{\delta \varphi})^2 \rangle +
\frac{\langle (\nabla \delta \varphi)^2\rangle}{2\,a^2} 
\ee
\noindent where we have also neglected the term with $ V''[\Phi_0]
\sim 3 \; H^2_0 \; \eta_V $ since  $ k^2/a^2 \gg H^2_0 $ for
subhorizon modes. Therefore, to leading order in slow roll we find
the renormalized expectation value of $T_{00s}$ is given by 
\be
\label{T00sfin} \langle T^0_{0s}\rangle_{ren} \simeq \frac{3
H^4_0}{(4 \; \pi)^2} \frac{\eta_V-4\,\epsilon_V}{\eta_V-3 \,
\epsilon_V} + \frac{1}{2}\Bigg\langle \dot{\delta \varphi}^2 +
\left(\frac{\nabla \delta \varphi}{C(\eta)}\right)^2
\Bigg\rangle_{ren} 
\ee To obtain the renormalized expectation
value in eq.(\ref{T00sfin}) one can set all slow roll parameters
to zero to leading order and simply consider a massless scalar
field minimally coupled in de Sitter space time
and  borrow the result from eq.(\ref{Tsiglea}). We find\cite{anom}
\be \langle
T^0_{0s} \rangle_{ren} =
 \frac{H^4_0}{(4 \; \pi)^2}\left[\frac{\eta_V-4\,\epsilon_V}{\eta_V-3 \,
\epsilon_V}-\frac{29}{30}+
\mathcal{O}(\epsilon_V,\eta_\sigma,\eta_V)\right]\label{T00sfinal2}
\ee
The last term in eq. (\ref{T00sfinal2}) is completely determined
by the {\bf trace anomaly} of a minimally coupled scalar field in de
Sitter space time\cite{fordbunch,BD,ddhf}

\subsection{Quantum Corrections to the Inflaton potential from
the Tensor perturbations}

Tensor perturbations correspond to massless fields with two physical
polarizations. The energy momentum tensor for gravitons only depends on
derivatives of the field $ h^i_j $ therefore its  expectation value in
the Bunch Davies (BD) vacuum does not feature infrared singularities in
the limit $\epsilon_V \rightarrow 0$. The absence of infrared
singularities in the limit of exact de Sitter space time, entails
that we can extract the leading contribution to the effective
potential from tensor perturbations by evaluating the expectation
value of $T_{00}$ in the BD vacuum in exact de Sitter
space time, namely by setting all slow roll parameters to zero. This
will yield the leading order in the slow roll expansion.

Because de Sitter space time is maximally symmetric, the expectation
value of the energy momentum tensor is given by\cite{BD}
\be \label{trace}
\langle T_{\mu \nu} \rangle_{BD} = \frac{g_{\mu
\nu}}{4} \; \langle T^{ \alpha}_{\alpha } \rangle_{BD}
\ee
\noindent and $ T^{ \alpha}_{\alpha }$ is a space-time constant,
therefore the energy momentum tensor is manifestly covariantly
conserved.  A large body of work has been
devoted to study the trace anomaly in de Sitter space time
implementing a variety of powerful covariant regularization methods that
preserve the symmetry\cite{ddhf,BD,fordbunch}
yielding a renormalized value of $ \langle T_{\mu \nu} \rangle_{BD} $
given by eq. (\ref{trace}). Therefore, the full
energy momentum tensor is completely determined by the {\bf  trace
anomaly}. The contribution to the trace anomaly from gravitons has been
given in refs.\cite{fordbunch,BD,ddhf}, it is 
\be \label{traza} 
\langle T^{\alpha}_{\alpha } \rangle_t = -\frac{717}{80 \; \pi^2} \;  H^4_0
\quad \mbox{and} \quad 
\langle T^0_{0} \rangle_t = -\frac{717}{320 \;  \pi^2}\; H^4_0\; .
\ee 

\subsection{Summary of Quantum Corrections to the Observable Spectra}

In summary, we find that the effective potential at one-loop is given by,
$$
\delta V(\Phi_0)  = \frac{H^4_0}{(4 \; \pi)^2} \Bigg[\frac{\eta_V-4 \;
\epsilon_V}{\eta_V-3 \; \epsilon_V}+ 
\frac{3\,\eta_\sigma}{\eta_\sigma-\epsilon_V}+
\mathcal{T}_{\Phi}+ \mathcal{T}_s+\mathcal{T}_t +\mathcal{T}_{\Psi}+
\mathcal{O}(\epsilon_V,\eta_V,\eta_\sigma,\mathcal{M}^2) \Bigg] \; ,
$$
\noindent where $(s,t,\sigma,\Psi)$ stand for the contributions
of the scalar metric, tensor fluctuations, light boson field $\sigma$ and light
fermion field $\Psi$, respectively. We have
\be
\mathcal{T}_{\Phi}  =  \mathcal{T}_s = -\frac{29}{30}  \quad ,  \quad 
\mathcal{T}_t  =  -\frac{717}{5} \quad ,  \quad 
\mathcal{T}_{\Psi}  =  \frac{11}{60} \label{Tt}
\ee
The terms that feature the \emph{ratios } of combinations of slow
roll parameters arise from the infrared or superhorizon
contribution from the scalar density perturbations and scalar
fields $\sigma$ respectively. The terms $\mathcal{T}_{s,t,\Psi}$
are completely determined by the trace anomalies of scalar,
graviton and fermion fields respectively. Writing $H^4_0 =
V(\Phi_0) \;  H^2_0/[3 \; M^2_{Pl}]$ we can finally write the effective
potential to leading order in slow roll
\be
V_{eff}(\Phi_0) = V(\Phi_0)\Bigg[1+ \frac{H^2_0}{3 \; (4 \; \pi)^2 \;
M^2_{Pl}}\Bigg( \frac{\eta_V-4\,\epsilon_V}{\eta_V-3 \; \epsilon_V}+
\frac{3\,\eta_\sigma}{ \eta_\sigma-\epsilon_V}-\frac{2903}{20}\Bigg)
\Bigg] \label{Veffin}
\ee
There are several  {\bf remarkable} aspects of this result:

i) the infrared
  enhancement as a result of the near scale invariance of scalar
  field fluctuations, both from scalar density perturbations as
  well as from a light scalar field, yield corrections of \emph{zeroth
  order in slow roll}. This is a consequence of the fact that
  during slow roll the particular combination $ \Delta_\sigma = \eta_\sigma-
\epsilon_V $ of slow roll parameters yield a natural infrared cutoff.

ii) the final one   loop contribution to the effective potential displays the
  effective field theory dimensionless parameter $H^2_0/M^2_{Pl}$.

iii) the last term is completely
  determined by the trace anomaly, a purely geometric result of the
short distance properties of the theory.

The quantum corrections to the effective potential lead to quantum corrections
to the amplitude of scalar and tensor fluctuations with the result\cite{anom}
\bea &&
|{\Delta}_{k,eff}^{(S)}|^2 = |{\Delta}_{k}^{(S)}|^2 \left\{ 1+
\frac23 \left(\frac{H_0}{4 \; \pi \; M_{Pl}}\right)^2
\left[1+\frac{\frac38 \; r \; (n_s - 1) + 2 \; \frac{dn_s}{d \ln
k}}{(n_s - 1)^2} + \frac{2903}{40} \right] \right\} \cr \cr &&
|{\Delta}_{k,eff}^{(T)}|^2 =|{\Delta}_{k}^{(T)}|^2  \left\{ 1
-\frac13 \left(\frac{H_0}{4 \; \pi \; M_{Pl}}\right)^2
\left[-1+\frac18 \; \frac{r}{n_s - 1}+ \frac{2903}{20}
\right]\right\} \; , \cr \cr &&
r_{eff} \equiv \frac{|{\Delta}_{k,eff}^{(T)}|^2}{|{\Delta}_{k,eff}^{(S)}|^2}
= r \;  \left\{ 1-\frac13 \left(\frac{H_0}{4 \; \pi \; M_{Pl}}\right)^2
\left[1+\frac{\frac38 \; r \; (n_s - 1) +
\frac{dn_s}{d \ln k}}{(n_s - 1)^2} + \frac{8709}{20} \right] \right\} \; .
\eea 
The quantum corrections turn out to {\bf
enhance} the scalar curvature fluctuations and to {\bf reduce} the
tensor fluctuations as well as their ratio $r$.

Acknowledgment: H J de V thanks the organizers of JGRG15 for their
kind hospitality in Tokyo.

\appendix

\section{One loop effective potential and equations of motion in
Minkowski space-time: a comparison}

In this appendix we establish contact with familiar effective
potential both at the level of  the equation of motion for the
expectation value of the scalar field, as well as the expectation
value of $T_{00}$.

In Minkowski space time the spatial Fourier transform of the field
operator is given by
\be \varphi_{\vk}(t) = \frac{1}{\sqrt{2\omega_k}}\left[ a_{\vk} \;
e^{-i\omega_k\,t} + a^{\dagger}_{-\vk} \;
e^{i\omega_k\,t}\right]\,, \ee
\noindent where the vacuum state is annihilated by $a_{\vk}$ and the
frequency is given by
\be
\omega_k = \sqrt{k^2+V^{''}(\Phi_0)} \,.
\ee
The one-loop contribution to the equation of motion (\ref{1lupeqn})
is given by
\be\label{1lupmink}
\frac{V^{'''}(\Phi_0)}{2} \; \langle
[\varphi(\vx,t)]^2\rangle = \frac{V^{'''}(\Phi_0)}{8  \, \pi^2}
\int^{\Lambda}_0 \frac{k^2}{\omega_k} \; dk = \frac{d}{d\Phi_0}
\left[ \frac{1}{4\pi^2} \int^{\Lambda}_0 k^2 \; \omega_k \; dk
\right] \,.
\ee
The expectation value of $T_{00}=\mathcal{H}$ (Hamiltonian density)
in Minkowski space time is given up to one loop by the following
expression
\be \langle T_{00} \rangle = \left\langle
\frac{1}{2} \; \dot{\phi}^2+\frac{1}{2} \; \left(\nabla \phi
\right)^2+V(\phi) \right\rangle =  \frac{1}{2} \; {\dot{\Phi_0}}^2 +
V(\Phi_0)+ \left\langle
\frac{1}{2} \; \dot{\varphi}^2+\frac{1}{2} \; \left({\nabla
\varphi}\right)^2+\frac{1}{2} \; V^{''}(\Phi_0) \; \varphi^2
+\cdots\right\rangle \;. \ee
The expectation value of the fluctuation contribution  is given by
\be \left\langle
\frac{1}{2} \; \dot{\varphi}^2+\frac{1}{2} \; \left({\nabla
\varphi}\right)^2+\frac{1}{2} \; V^{''}(\Phi_0) \; \varphi^2
+\cdots\right\rangle = \frac{1}{4\pi^2} \int^{\Lambda}_0
k^2  \; \omega_k \; dk  =
\frac{\Lambda^4}{16\pi^2}+\frac{V^{''}(\Phi_0) \; \Lambda^2}{16 \,\pi^2}-
\frac{[V^{''}(\Phi_0)]^2}{64 \, \pi^2}
\ln\frac{4 \,\Lambda^2}{V^{''}(\Phi_0)} \;.
\ee
Renormalization proceeds as usual by writing the bare Lagrangian in
terms of the renormalized potential and counterterms. Choosing the
 counterterms to cancel the quartic,
quadratic and logarithmic ultraviolet divergences, we obtain the
familiar renormalized one loop effective potential
\be \label{potefM}
V_{eff}(\Phi_0) = V_R(\Phi_0)+\frac{[V^{''}_R(\Phi_0)]^2}{64 \, \pi^2}
\ln\frac{V^{''}_R(\Phi_0)}{M^2} \; ,
\ee
\noindent where $M$ is a renormalization scale.  Furthermore from
eq. (\ref{1lupmink}) it is clear that the equation of motion for
the expectation value is given by
\be\label{eqnmink}
\ddot{\Phi}_0+ V^{'}_{eff}(\Phi_0)= 0 \;.
\ee

\end{document}